\documentclass{article}
\usepackage{frascatiphys}
\usepackage{graphicx}
\usepackage{color}
\definecolor{yellow}{rgb}{0.95,0.75,0.1}
\definecolor{orange}{rgb}{0.95,0.4,0.1}
\definecolor{red}{rgb}{1,0,0}
\definecolor{green}{rgb}{0,1,0}
\definecolor{blue}{rgb}{0,0.5,1}

\definecolor{lblue}{rgb}{0,0.8,1}
\definecolor{dblue}{rgb}{0,0,1}
\definecolor{dgreen}{rgb}{0,0.65,0}
\definecolor{lila}{rgb}{0.8,0,0.8}
\definecolor{violet}{rgb}{1,0,0.9}
\definecolor{grey}{rgb}{0.3,0.3,0.3}

\definecolor{contoura}{rgb}{0,0,1}
\definecolor{contourb}{rgb}{0,1,1}
\definecolor{contourc}{rgb}{0,1,0}
\definecolor{contourd}{rgb}{0.95,0.75,0.1}
\definecolor{contoure}{rgb}{1,0,0}
\definecolor{contourf}{rgb}{1,0,1}

\newcommand {\black} {\color{black}}

\newcommand {\dblue} {\color{dblue}}

\newcommand {\dgreen} {\color{dgreen}}

\newcommand {\red} {\color{red}}
\newcommand {\orange} {\color{orange}}

\def\ARAA{{ Annual Rev. of Astron. \& Astrophys.} }
\def\ApJ{{ Astrophys. J.} }
\def\ApJL{{ Astrophys. J. Lett.} }

\def\ApP{{ Astropart. Phys.} }
\def\AA{{ Astron. \& Astroph.} }

\def\AAL{{ Astron. \& Astroph. Lett.} }

\def\JCAP{{ J. of Cosm. \& Astrop. Phys.} }

\def\MNRAS{{ Month. Not. Roy. Astr. Soc.} }
\def\Nature{{ Nature} }

\def\PASP{{ Publ. Astron. Soc. Pac.}Ê}

\def\PRD{{ Phys. Rev.} {\bf D} }
\def\PRL{{ Phys. Rev. Lett.} }
\def\RMP{{ Rev. Mod. Phys.} }
\def\RPP{{ Rep. Pro.Phys.} }
\def\Science{{ Science}Ê}

\def\simle{\lower 2pt \hbox {$\buildrel < \over {\scriptstyle \sim }$}}
\def\simge{\lower 2pt \hbox {$\buildrel > \over {\scriptstyle \sim }$}}

\begin{document}
\title{ THE NATURE AND ORIGIN OF ULTRA-HIGH ENERGY COSMIC RAY PARTICLES
}
\author{{{Peter L. Biermann$^{1,2,3,4}$}}, Lauren\c{t}iu  I. Caramete$^{5,1}$, \\ Federico Fraschetti$^{6}$, L{\'a}szl{\'o} {\'A}. Gergely$^{7}$,\\ Benjamin C. Harms$^{3}$, Emma Kun$^{7}$, Jon Paul Lundquist$^{8}$, \\ Athina Meli$^{9}$, Biman B. Nath$^{10}$, Eun-Suk Seo$^{11}$, \\ Todor Stanev$^{12}$, and Julia Becker Tjus$^{13}$.}
%Peter L. Biermann\footnote{\em MPI for Radioastr., Bonn, Germany}, \footnote{\em Dept. %of Phys., Karlsruhe Inst. for Tech. KIT}, \footnote{\em  Dept. of Ph. \& A., U. Alabama, %Tuscaloosa, AL, USA}, \footnote{\em Dept. of Phys. \& Astron., U. Bonn, Germany},\\ 
%Lauren\c{t}iu  I. Caramete\footnote{\em Inst. Spac. Sci., Bucharest, Romania},\\ 
%Federico Fraschetti\footnote{\em Dept. Phys. \& Astron., Univ. Arizona, Tucson, USA},\\ %L{\'a}szl{\'o} {\'A}. Gergely\footnote{\em Inst. of Phys., Univ. Szeged, Hungary},\\
%Benjamin C. Harms\footnote{\em Dept. of Ph. \& A., U. Alabama, Tuscaloosa, AL, USA},\\ %Emma Kun\footnote{\em Inst. of Phys., Univ. Szeged, Hungary},\\ 
%Jon Paul Lundquist\footnote{\em Dept. of Phys., Univ. of Utah, Salt Lake City, USA},\\ %Athina Meli\footnote{\em Dept. of Phys., Univ. Gent, Belgium},\\ 
%Biman B. Nath\footnote{\em Raman Res. Res. Inst., Bangalore, India},\\ 
%Eun-Suk Seo\footnote{\em Dept. of Phys., Univ. Maryland, USA},\\ 
%Todor Stanev\footnote{\em Bartol Res. Inst., Univ. Delaware, USA},\\ 
%and Julia Becker Tjus\footnote{\em Dept. of Phys., Univ. Bochum, Germany}
% %\end{center}

%\vskip0.5cm
%\eject
%\centerline{Jul 29, 2016; review after lectures at Vulcano/Nijmegen}
%\vskip0.5cm

\maketitle
\baselineskip=11.6pt
\begin{abstract}
We outline two concepts to explain Ultra High Energy Cosmic Rays (UHECRs), one based on radio galaxies and their relativistic jets and terminal hot spots, and one based on relativistic Super-Novae (SNe) or Gamma Ray Bursts (GRBs) in starburst galaxies, one matching the arrival direction data in the South (the radio galaxy Cen A) and one in the North (the starburst galaxy M82).  The most likely identification of the origin of observed Gravitational Wave (GW) events is stellar binary black hole (BH) mergers in starburst galaxies such as M82 with the highest rate of star formation, so the highest far-infrared (FIR) luminosity, at the edge of the universe visible in 10 - 300 Hz GWs; at low heavy element abundance $Z_{ch}$ the formation of stellar BHs extends to a larger mass range.  A radio galaxy such as Cen A sequence of events involves first the merger of two Super-Massive Black Holes (SMBHs), with the associated ejection of low frequency GWs, then the formation of a new relativistic jet aiming into a new direction: ubiquitous neutrino emission follows  accompanied by compact TeV photon emission, detectable more easily if the direction is towards Earth. The ejection of UHECRs is last.  Both these sites are the perfect high energy physics laboratory:    We have observed particles up to ZeV, neutrinos up to PeV, photons up to TeV, 30 - 300 Hz GW events, and hope to detect soon of order $\mu$Hz to mHz GW events.  Energy turnover in single low frequency GW events may be of order $\sim 10^{63}$ erg.  How can we further test these concepts?  First of all by associating individual UHECR events, or directional groups of events, with chemical composition in both the Telescope Array (TA) Coll. and the Auger Coll. data.  Second by identifying more TeV to PeV neutrinos with recent SMBH mergers.  Third by detecting the order $<$ mHz GW events of SMBH binaries, and identifying the galaxies host to the stellar BH mergers and their GW events in the range up to 300 Hz.  Fourth by finally detecting the formation of the first generation of SMBHs and their mergers, surely a spectacular discovery.  \footnote{Institutions:  $^{1}$ {\em MPI for Radioastr., Bonn, Germany};
$^{2}$ {\em Dept. of Phys., Karlsruhe Inst. for Tech. KIT};
$^{3}$ {\em Dept. of Phys. \& Astron., U. Alabama, Tuscaloosa, AL, USA}; 
$^{4}$ {\em Dept. of Phys. \& Astron., U. Bonn, Germany};
$^{5}$ {\em Inst. Spac. Sci., Bucharest, Romania};
$^{6}$ {\em Dept. Phys. \& Astron., Univ. Arizona, Tucson, USA};
$^{7}$ {\em Inst. of Phys., Univ. Szeged, Hungary}; 
$^{8}$ {\em Dept. of Phys., Univ. of Utah, Salt Lake City, USA}; 
$^{9}$ {\em Dept. of Phys., Univ. Gent, Belgium};
$^{10}$ {\em Raman Res. Res. Inst., Bangalore, India};
$^{11}$ {\em Dept. of Phys., Univ. Maryland, USA};
$^{12}$ {\em Bartol Res. Inst., Univ. Delaware, USA};
$^{13}$ {\em Dept. of Phys., Univ. Bochum, Germany};
with help by Heino Falcke (Radboud Univ., Nijmegen, Netherlands),
Sera Markoff (Univ. Amsterdam, Netherlands),
I. Felix Mirabel (Buenos Aires, Argentina; Paris, France), 
and Vitor de Souza (Univ. Sao Paolo at San Carlos, Brazil).}
\end{abstract}
%\footnote{Coauthors:  Lauren\c{t}iu  I. Caramete, 
%{\em Inst. Spac. Sci., Bucharest, Romania};
%Federico Fraschetti,
%{\em Dept. Phys. \& Astron., Univ. Arizona, Tucson, USA};
%L{\'a}szl{\'o} {\'A}. Gergely,
%{\em  Inst. of Phys., Univ. Szeged, Hungary};
%Benjamin C. Harms,
%{\em Dept. of Ph. \& A., U. Alabama, Tuscaloosa, AL, USA};
%Emma Kun,
%{\em Inst. of Phys., Univ. Szeged, Hungary};
%Jon Paul Lundquist,
%{\em Dept. of Phys., Univ. of Utah, Salt Lake City, USA};
%Athina Meli,
%{\em Dept. of Phys., Univ. Gent, Belgium};
%Biman B. Nath,
%{\em Raman Res. Res. Inst., Bangalore, India};
%Eun-Suk Seo,
%{\em Dept. of Phys., Univ. Maryland, USA};
%Todor Stanev,
%{\em Bartol Res. Inst., Univ. Delaware, USA};
%Julia Becker Tjus,
%{\em Dept. of Phys., Univ. Bochum, Germany}}
%
\baselineskip=14pt
%
%\footnote{with help by Heino Falcke (Radboud Univ., Nijmegen, Netherlands),
%Sera Markoff (Univ. Amsterdam, Netherlands), I. Felix Mirabel (Buenos Aires, %Argentina; Paris, France), and Vitor de Souza (Univ. Sao Paolo at San Carlos, Brazil)}

\section{Introduction:  Challenges of High Energy Events}

Today we have an abundance of riches, with almost certainly more to come:  We have a very well defined spectrum of ultra high energy cosmic ray (UHECR) particles, with chemical composition information, and a kink up in the overall spectrum, near $3 \cdot 10^{18}$ eV.  We have directional information, with a weak directional hot spot in the South, and a much better defined directional hot spot in the North, both suggesting specific galaxies as sources, that have long been ranked as the leading candidates of their activity in the local universe: The radio galaxy Cen A in the South, with a recent SMBH binary merger, and the starburst galaxy M82 in the North with relativistic Super-Novae (SNe) or Gamma Ray Bursts (GRBs). Starburst galaxies just as radio galaxies can produce both energetic particle populations up to order $10^{21} \, Z$ eV, where $Z$ is the charge of the nucleus.

We also have very high energy neutrino events, that may originate in recent mergers of Super-Massive Black Holes (SMBHs): How can we recognize such events?  In a merger between two galaxies, both with a super-massive black hole at the center, orbital angular momentum wins, and so leads to a new direction of the final spin.  The final spin direction defines the axis of the new relativistic jet, which has to plow a new channel through the dense material near the newly defined center of the merged galaxy.  This plowing leads to powerful injection, acceleration and particle interaction, therefore giving rise to lots of high energy neutrinos, detectable in the case the new jet points at Earth, and then recognizable via a flat spectrum to near THz radio frequencies.  Thus, these mergers give rise to a low frequency GW background, yet to be discovered, in the range from order 1 mHz on down.

Furthermore, we now have the detections of GW Events from the merging of stellar black holes.  This reminds us that massive stars almost all are in comparable mass binary star systems, and so the GW event rate ought to scale directly with the supernova rate of massive stars detectable in their radio emission.  The supernova rate in turn scales with the FIR luminosity.  Some of all supernovae also explode as relativistic SNe or GRBs, allowing yet higher energy particles to be produced.  And so any starburst galaxy produces in parallel SNe, GRBs, and as a consequence particles to $10^{17.3} \, Z$ eV from SNe,  up to order $ 10^{21} \, Z$ eV from relativistic SNe or GRBs, the corresponding high energy neutrinos, and GW events in the 10 - 300 Hz range.

\section{The spectrum of cosmic rays}

In combination of Auger Coll. \cite{Aab15} and the Telescope Array Coll. (TA) \cite{Abbasi16} data with other experiments we now have a consistent spectrum of cosmic rays, that readily finds an interpretation \cite{Thoudam16}; for recent reviews see \cite{Letessier-Selvon11,Kotera11}, and fundamental books \cite{Stanev10,Gaisser16}. This explanation requires a contribution from massive stars such as Wolf-Rayet stars, which have powerful stellar winds.  In this context it is of interest to note that we have good radio data on supernova explosions racing through a wind \cite{Brunthaler10,Fransson98,Soderberg05,Soderberg06,Soderberg08,Soderberg10} 
\cite{Krauss11,Milisavljevic13,Kimani16,DeWitt16}. These early data, covering six explosions that probably were Wolf-Rayet stars, so Blue Supergiant stars, and two explosions that probably were a Red Supergiant star, all suggest the same: i)  The speed of the shock is initially about 0.1 c, ii)  the magnetic field in the post-shock region is about 1 Gauss at a radial distance of $10^{16}$ cm, iii) the run of the magnetic field is close to $r^{-1}$ with $r$ being the radial coordinate.  The numbers allow the two limits of the energy that can be reached to be worked out, i)  in the limit that the magnetic field is mainly parallel to the shock normal, the Bohm limit \cite{Drury83}; and conversely ii) in the limit that the magnetic field is mainly parallel to the shock surface, the Jokipii limit \cite{Jokipii87}:  These numbers suggest $E_{Bohm} \; = \; 10^{15.3 \pm 0.3} \, Z$ eV; and $E_{Jokipii} \; = \; 10^{17.3 \pm 0.2} \, Z$ eV, where $Z$ is the charge of the particle.  It is striking that the first energy corresponds well to the knee, and the second energy corresponds well to the ankle, suggesting that both scattering regimes are important, Bohm and Jokipii.  The very fact, that for the two types of stars, with very different wind velocities, and so different wind densities, give rise to the same magnetic field, precludes the Bell-Lucek mechanism \cite{LucekBell00,BellLucek01}  as the key path to attain these magnetic fields in the observed shock.  The Bell-Lucek mechanism may have worked prior to the SN-explosion, when very often strong eruptive events are detected \cite{Svirski14,GalYam14,Ofek14,Strotjohann15,Tartaglia16}, as also observed for $\eta$ Carinae in the 19th century.  We note that the main sequence stages of these stars do have a magnetic field, but usually of order only a few hundred Gauss \cite{Hubrig14,Kholtygin15,Martins10,Wade11,Wade16}.  However, as these stars had an inner convective zone with magnetic fields attaining the $10^{6}$ or even $10^{7}$ Gauss range \cite{Biermann93II} and the surface of the star due to the strong winds is reaching deep down, it is also possible that these magnetic fields seen in the explosion were derived from the former inner magnetic fields.  That would be consistent with their properties being so similar in the cases well observed sofar.

Relativistic SNe or GRBs would obviously enhance the particle energies further, probably reaching $10^{21} \, Z$ eV; however, the IceCube Coll. evidence \cite{Aartsen16} speaks against a dominant GRB contribution of the UHECR flux.  On the other hand, the TA Coll. evidence suggests a proton contribution from the starburst galaxy M82, see below.

\section{Distribution of arrival directions: all sky}

The Auger Coll. finds a weak correlation in arrival direction of events with the radio galaxy Cen A, consistent with a very old prediction \cite{Ginzburg63a}.

The TA Coll. finds a much more pronounced directional ``hot spot" in arrival directions not far from the starburst galaxy M82, consistent with the expectation that relativistic SNe or GRBs are a special form of massive star explosions, that lead to the acceleration of protons and nuclei to energies near $10^{21}$ eV (e.g. \cite{Biermann94a,Biermann94b}); for a quantitative approach see \cite{Milgrom95,Vietri95,Waxman95}, with many later papers and reviews, e.g. \cite{Piran99,Piran04,Ioka10,Meszaros13}.  We note that close to 100 percent of all massive stars are in binary systems of comparable mass \cite{Chini12}, allowing mass transfer between the two stars and also the formation of two stellar BHs \cite{Heger03}.

The spread of arrival directions (of order 10 degrees) as well as the central shift in direction (near 0 degrees for Cen A, and near 20 degrees for M82) can be understood as scattering and orbit bending in the Galactic wind of our Galaxy, subject to a $k^{-2}$ turbulence, with $k$ the wavenumber, indicative of what can be described as compressible turbulence,  super-sonic and/or super-Alfv{\'e}nic turbulence, or shock-dominated irregularities \cite{Kolmogorov41a,Kolmogorov41b,Federrath13,Biermann15}.  A detailed exploration of the implications (see, e.g., the Faraday sky in \cite{Oppermann12}), and the evidence for a Galactic wind \cite{Everett08,Everett12}, and \cite{Uhlig12} has yet to be done.

How can we check on such interpretations?  In the case of a starburst galaxy such as M82 the magnetic field of a Galactic wind, even when highly irregular, ought to impose a certain geometric pattern on the arrival directions, strongly depending on the charge of the particles, since all bending runs with energy over charge $E/Z$; there is now some evidence of this geometry.  Additionally, along with massive star SNe there ought to be occasional events of GWs, such as observed already {\cite{Abbott16a,Abbott16b}.  The rate of massive star SNe as well as GW events ought to scale with the rate of massive star formation, so the luminosity of a starburst in the FIR \cite{Biermann77,Kronberg85,Sanders96,Lagache05}.  The higher the rate, the more likely it is to detect an event  within a certain time window; on the other hand, with a luminosity function decreasing with luminosity \cite{Lagache05} we have the highest such rate near the limit of sensitivity, when the lowest detectable flux corresponds to the most powerful starburst within the survey volume.  Lagache et al. state ``Luminosity function evolution is such that the power output is dominated by LIRGs at $z \, \simeq \,  0.7$ (although they represent only 3\% of the galaxies) and by ULIRGs at $z \, \simeq \, 2.5$ (although they represent only 1\% of the galaxies).", so that the brightest galaxies will dominate in sampling GW events; here LIRG means ``Luminous Infra-Red Galaxy", with a star formation rate of about $10 \; - \; 100 \, {\rm M_{\odot} \, yr^{-1}}$, and ULIRG means ``Ultra-Luminous Infra-Red Galaxy", with a yet higher star formation rate and possibly additional feeding from an Active Galactic Nucleus (AGN).  However, if relativistic SNe or GRBs and GW events depend on the formation of Black Holes, for which the mass range depends on the heavy element abundance $Z_{ch}$ \cite{Heger03}, then all this may get more subtle, and will require more exploration.  To be compatible with the IceCube Coll. limits \cite{Aartsen16}, it appears as a testable concept that the directional hot spot due to M82 is only a localized patch of very energetic protons, on top of a $4 \pi$ spread of nuclei from the radio galaxy Cen A; this may explain why the TA Coll. flux tends to be a tad above the Auger Coll. flux.

In the case of radio galaxies it is of interest to focus on merger events of super-massive binary black holes \cite{Kun16}, which generally give rise to a new spin direction of the SMBH after the merger: in that case the jet has to plow a new channel, maximizing injection, acceleration, and interaction \cite{Toomre72,Nellen93,Gopalkrishna03}, \cite{Gergely09,Becker09,Gergely10,Gopalkrishna12,Tapai13,Tjus14} so providing a prime site for the acceleration of UHECRs: Kun et al. argue that in the case where after the merger the relativistic jet points at Earth this stage can be identified by observing a flat radio spectrum from GHz frequencies all the way to the FIR, so near THz radio frequencies, such as detected by IRAS, WMAP or PLANCK.  A first search identified two neutrino track events (with with about a degree in directional uncertainty: \cite{Aartsen14,Aartsen15}) with such radio sources, with a very low combined probability that the identification is random. The task remains to identify more track events.  This interpretation implies that recent SMBH mergers are prodigious producers of UHECRs.  It also implies that there ought to be a strong background of gravitational waves in the range order 1 $\mu$Hz to 1 mHz \cite{Mignard12,Shannon15,Lentati16,Verbiest16}, allowing for the main part of the spectrum of SMBHs, between $\sim$$10^{6}$ and $\sim$$10^{8} \, {\rm M_{\odot}}$, and a possibly relevant redshift range between 10 and 100 \cite{Biermann06,Biermann14}.  The observation of such slow GW events will remain a task for the future, while the Lisa Path Finder mission allows grounds for optimism \cite{Armano16} in the long term.

\section{The spectrum below $3 \cdot 10^{18}$ eV, the ankle}

LOFAR \cite{Buitink16} and Kaskade-Grande \cite{Arteaga15} have demonstrated that the long expected \cite{Rachen93a,Rachen93b} extragalactic proton component may be there at these lower energies.  One unanswered question here is whether this flux results from a long time integral of the locally produced high energy cosmic ray particles or is the long distance accumulation from various active sources \cite{Liu16}.  This distinction depends on the probably extremely inhomogeneous structure of the large scale intergalactic magnetic field.  This magnetic field is embedded into the accretion flow towards filaments and sheets, and so there ought to be a specific energy beyond which particles can no longer escape this accretion flow and just propagate along filaments and sheets.  This has focussing  as consequence, along filaments the flux of particles caught does not diminish with distance except by scattering to higher energy and ensuing escape, and along sheets the flux weakens with inverse distance \cite{Ryu98}.  This could produce a strong spectral turn-down feature, which is not seen except at a characteristic energy of about $6 \cdot 10^{19}$ eV.  This turn-down could also be due to spatial limits to acceleration in the source, or to interaction with the microwave of far-infrared background.  We see no such feature at lower energy within the current data.

\section{Jets and terminal hot spots in relativistic jets}

Shocks in relativistic flow appear as prime candidates to inject and accelerate protons and heavier nuclei to ultra high energy; remember, that GRBs are commonly interpreted as relativistic jets with initially very high Lorentz factor (to several hundred), while relativistic jets emanating from SMBHs may range from Lorentz factor barely above unity to about 100 (see, e.g., \cite{Gopalkrishna10}); relativistic SNe are one other possibility \cite{Ioka10}.  These shocks may plow through a starburst region with the standard galactic cosmic ray spectrum and use them for injection to ultra high energy \cite{Gopalkrishna10,Biermann12,Todero15}.

Here we briefly review the relevant arguments based on the series of papers started by \cite{Falcke95a,Falcke95b}, and \cite{Markoff03}, based on some ideas in \cite{Biermann87}, and used again in \cite{Biermann11,Biermann14,Biermann15}:

First some comments on jets:  Jet flow suffers dramatically from adiabatic losses, and yet keeps going from near the central BH to $\simle$ 3,000 $r_{S}$ to order $10^{24}$ cm, sometimes to $10^{25}$ cm or even more, where $r_S$ is the Schwarzschild radius of the central BH.  Jets can be expected to start at a near-relativistic speed of sound, but cool down rapidly.  Each shock system consumes only a minute fraction of kinetic energy (remember, entropy is increased in any shock).  The observational evidence suggests a spiral (almost DNA-like geometry) pattern of highly oblique shocks: so we have continuous re-acceleration (see, also, \cite{Zdziarski16}) of the particle population \cite{Meli13}. However, each shock is strong suggesting  that the internal sound-speed is sub- or only weakly relativistic.  

The ubiquitous cutoff spectrum of non-thermal emission near $3 \cdot 10^{14}$ Hz observed \cite{Rieke76,Rieke79,Rieke82,Rieke80,Bregman81,Stocke81, Sitko83} 
\cite{Brodie83,Roeser86,Meisenheimer86,Perez88} since the mid-seventies in jets, terminal hot spots and compact unresolved Active Galactic Nuclei can be explained as the combined effect of first protons (or nuclei) getting accelerated to the synchrotron loss limit, and giving rise to a $E^{-2}$ spectrum.  Such a spectrum results in a $k^{-5/3}$ spectrum of magnetic irregularities (the same spectrum as \cite{Kolmogorov41a,Kolmogorov41b}, but here via excitation at all wavelengths, \cite{Bell78a,Bell78b}). Note that nearly every shock is preceded by another shock with injection of both turbulence and energetic particles further upstream \cite{Meli13}.  Electrons then get accelerated in that same spectrum of irregularities again to their loss limit, and at their maximal energy give a maximal synchrotron frequency independent of all parameters,
$\nu^{\star}_e \; \simle \; 3 \cdot 10^{14} \, {\rm Hz}$.  Generalizing for synchrotron losses and photon interaction losses is straight-forward and does not modify the numbers substantially.  This translates to a maximal energy for protons of $E_{p,max}  \simeq \, 1.4 \cdot 10^{20} {\rm eV} \,\{{\nu^{\star}_e}/{ 3 \cdot 10^{14} \, {\rm Hz}}\}^{1/2} \, B^{-1/2} $, where $B$ is the magnetic field in Gauss, typically observed to be of order mGauss in compact nuclei.  There is also a spatial limit $10^{21} \, L_{46}^{1/2} \; {\rm  eV}$, where $L_{46}$ is the jet power in units of $10^{46} \, {\rm erg/s}$  \cite{Lovelace76,Falcke95b}; we note that the maximal magnetic field in jets is given by a Poynting flux jet which scales with the square of the magnetic field.  Therefore we have in combination $E_{p,max}  \simeq \, 1.4 \cdot 10^{21} {\rm eV} $  (no boosting assumed here).  Thus UHECR particles are required to explain why the feature is so ubiquitous.

Combining then the maximal emission frequencies of protons and electrons we find
$\{{\nu_{syn, p, max}}/{\nu_{syn, e, max}}\} \, = \, \{{m_p}/{m_e}\}^3 $, matching the first characteristic of the double-bump spectrum of blazars \cite{Biermann11}, the spectral distance between the two bumps.  Integrating then downstream (see also \cite{Biermann87,Sikora09,Zdziarski15}) from each shock along a stream-line following \cite{Kardashev62} gives \\ $\{{L_p}/{L_e}\} \; \sim \; \{{n_{p, 0} m_p}/{n_{e, 0} m_e}\} \, \{{\gamma_{p, max}}/{\ln(\gamma_{e,max} / \gamma_{e,min})}\} \, \{{m_{e}}/{m_{p}}\}^{+3} \; \sim \;  1 $ matching the the second characteristic of the double-bump spectrum of blazars, the crudely equal luminosity of the two bumps (e.g. \cite{Giommi12,Sol13}).  In this approach the blazar sequence arises from the dependencies on SMBH mass and boosting factor \cite{Biermann11}.  We propose this is the basic explanation of this observation.  Quite obviously, many interactions such as Inverse Compton of these two photon-bumps ensue and modify what we observe.  This requires proton (or nuclei) acceleration in the sources, in radio galaxies: after all, radio galaxies, perchance pointing at us,  are blazars.

One important test is the variability time-scale $\tau$:  $\tau \simeq r/(2 \, \Gamma^2 \, c)$ \cite{Piran99}, where $r$ is the radial distance, and $\Gamma$ is the Lorentz factor of the jet; minutes seen in TeV photons \cite{Aharonian07} imply $10^{16.6}$ cm, $\simeq$ 3000 $r_S$ of a $10^8 \; M_{\odot}$ SMBH for $\Gamma \simeq 100$.  This is near where jets turn into a conical outflow \cite{Yuan02,Marscher02,Markoff03}.  

This is an old prediction, but a new argument: Radiogalaxies are sources at energies $> \; 10^{20} $ eV, with an energy that may approach $\sim 10^{63}$ erg, possibly on occasion even more. Lower energy budget CR-sources are intergalactic shocks \cite{Hong14}, Gamma Ray Bursts \cite{Piran04}, micro-quasars \cite{Heinz02,Mirabel06}, jet-supernovae, pulsar wind nebulae \cite{Fang12}, powerful supernovae \cite{Biermann03,Ioka10}, and probably yet other activities we do not understand yet.  Ginzburg \& Syrovatskij \cite{Ginzburg63a} identified the nearby candidates before the discovery of UHECRs: the radio galaxies i) Cen A (= NGC 5128), a recent SMBH merger; ii) Vir A (= M87 = NGC 4486), a recent SMBH merger; and iii) For A (= NGC 1316),  perhaps also a SMBH merger, but in this case the radio morphology is ambiguous, while for Cen A and M87 the old jet directions are clearly visible, so a re-orientation of the dominant jet is recognizable, a direct consequence of the merger of two SMBHs.  Considering the energetics of the observed compact jets in radio galaxies, a hierarchical ranking can be derived of what radio galaxies may contribute \cite{Caramete16}, and the first three radio galaxies are just those already identified by \cite{Ginzburg63a}; the detailed statistics show, that Cen A is expected to dominate the entire integral of the UHECR contribution from radio galaxies lower down in the ranks.

\subsection{How do SMBHs start?}

The observational evidence suggests that the SMBH mass function \cite{Caramete10} starts around $3 \cdot 10^{6} \, {\rm M_{\odot}}$ and its shape can be fully explained by merging in the gravitational focussing limit \cite{Silk79,Gergely12}, minimizing any electromagnetic output \cite{Ensslin98,Caramete10}.

Why would star formation pick such a mass? First massive stars can form in dense groups in the gravitation of the Dark Matter potential well of a dwarf galaxy \cite{Strigari08,Donato09}: then stars agglomerate \cite{Spitzer69,Sanders70} to form a more massive star.  Massive stars also have winds, driven by radiation interaction with heavy elements (\cite{Lucy70} and many later papers): So their maximum mass attainable is several hundred $M_{\odot}$ at most \cite{Yungelson08}.  However, at zero heavy element abundance massive stars can grow to much higher mass, close to $10^{6} \, {\rm M_{\odot}}$.  At that point massive stars hit an instability, combining radiation pressure with subtle effects of General Relativity \cite{Appenzeller72a,Appenzeller72b}.  As a consequence they explode: So with infall their BH mass may reach  about $3 \cdot 10^{6} \; M_{\odot}$ explaining the observations.  A key prediction of this specific model is that the initial formation of the first generation of SMBHs is only allowed at metal abundance $Z_{ch}$ near zero, a test which we hopefully can make in the future.

There are alternate pictures (e.g. \cite{Munyaneza05,Munyaneza06}) using a specific  model of Dark Matter.  Other models just using massive stars and accretion would produce a large range in redshifts of when these SMBHs begin (e.g. \cite{Menci16}), as would a concept in which gas collapses directly to a compact object \cite{Spitzer67}, possibly a black hole.

The sky distribution of SMBHs allows further constraints on the origin to be set:  it can be shown in a graph, where colors are distance: Black, Blue, Green, Orange, Red, for the redshifts intervals in steps of 0.005 to 0.025, i.e. distance intervals of 20, to 100 Mpc.  

%The sky plot % 26 september 2007 %%%%% 2micron 5,978 black hole color coded
\begin{figure}[htb]
\begin{center}
\includegraphics[bb=0cm 0cm 21.0cm 29.7cm,viewport=1cm 1.0cm 21.0cm
11.5cm,clip,scale=0.65]{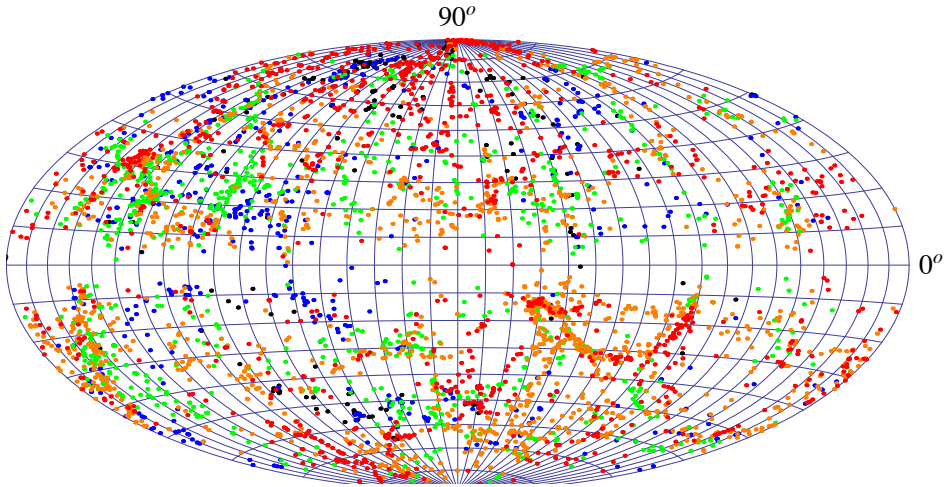}
\caption{The sky in super-massive black holes $> \, 3 \cdot 10^{7} \, {\rm M_{\odot}}$,  where colors are distance: {\black Black}, {\dblue Blue}, {\dgreen Green}, {\orange Orange}, {\red Red}, for the five redshifts intervals in steps of 0.005 to 0.025. The coordinate system is with the Galactic Plane across the center, and Galactic Center (GC) at the right/left edge (See \cite{Caramete10,Caramete11}).}
\end{center}
\label{SkyinBHs}
\end{figure}

The striking feature in this sky distribution is the semi-circular feature of hundreds of SMBHs, many of which are above $10^{8} \, M_{\odot}$: This can be understood as a consequence of the freeze-out of the expansion of a spherical disturbance \cite{Bashinsky01,Bashinsky02}, as discussed elsewhere (lectures by P.L. Biermann at the Chalonge meetings in Paris 2015 and 2016, and ensuing discussions).  Geometrically this could be due to the cut of an expanding baryonic spherical shell through a Dark Matter Zeldovich pancake \cite{Zeldovich70}, using the run-away cooling mechanism proposed in \cite{Biermann06} in shocks due to supersonic flow \cite{Fialkov14}; in this specific case the arc would correspond to the third bump in the Micro-Wave Back-Ground (MWBG) power spectrum.  This may require a very high redshift to get started.  A test of any such picture would be the common detection of such partial circular arcs of SMBHs; considering the skymap shown this may indeed be a common occurrence, sometimes corresponding to the first peak in the MWBG power spectrum.

\subsection{SMBH energetics}

As shown in work by P.P. Kronberg (lecture at DRAO Nov 2015) allowing for $P \, dV$-work in understanding the scale of energy of giant radio galaxies the total energy may reach rather close to a good fraction of $M_{SMBH} \, c^2$, allowing for other channels than just radio emission, possibly as close as $ \sim \, 1/2$ \cite{Hawking71}.  The two observed GW stellar BH merger events correspond to about 0.05 of $M \, c^2$ in GWs emitted.  One can speculate that stellar BHs ought to merge starting from a small spin, but that SMBHs in radio galaxies may start from near maximum spin, as it is derived from the orbital angular momentum of the merging SMBHs (mass and spin enter the maximally allowed efficiency).  It follows that there ought to be a powerful GW background due to the formation and merging of SMBHs, in the range between order 1 $\mu$Hz and order 1 mHz, depending on the exact mass range and redshift of the first generation of formation of SMBHs and their merging history \cite{Mignard12,Shannon15,Lentati16,Verbiest16}.  As shown in \cite{Biermann06} this redshift could be quite high.

We note that the existing observational limits of any GW background all pertain to either today, the recombination redshift \cite{Abbott09}, or earlier epochs even.  For any redshift $<$ 100, the maximum redshift allowed by the mechanism in \cite{Biermann06} there is no limit at all today at the frequency range given by SMBHs other than the observed energy density of Dark Energy (DE).

\section{Summary}

We have outlined two concepts to explain UHECRs, one based on radio galaxies and their relativistic jets and terminal hot spots, and one based on relativistic SNe or GRBs in starburst galaxies, one matching the arrival direction data in the South (the radio galaxy Cen A) and one in the North (the starburst galaxy M82).  The most likely identification of the origin of observed GW events is starburst galaxies such as M82 with the highest rate of star formation, so the highest FIR luminosity, at the edge of the universe visible in 10 - 300 Hz GWs; the value of the heavy element abundance $Z_{ch}$ restricts the mass range for stellar BHs \cite{Heger03}.  The radio galaxy sequence of events involves first the merger of two SMBHs, with the associated burst of low frequency GWs, then the formation of a new jet aiming into a new direction: ubiquitous neutrino emission follows (detectable more easily if the direction is towards Earth) accompanied by compact TeV photon emission. The ejection of UHECRs is last.

So these sites are the perfect high energy physics laboratory:    We have particles up to ZeV, neutrinos up to PeV, photons up to TeV, and of order $\mu$Hz to 300 Hz GW events; inside the source the energies may go higher.  Energy turnover in GW single events may approach $\sim 10^{63}$ erg, possibly on occasion even more.

How can we further test these concepts?  First of all by associating individual UHECR events, or directional groups of events, with chemical composition in both the TA Coll. and the Auger Coll. data.  Second by identifying more TeV to PeV neutrinos with recent SMBH mergers.  Third by detecting the order $\mu$Hz GW events and identifying the galaxies host to the stellar BH mergers and their GW events in the range up to $\sim$ 300 Hz.  Fourth by finally detecting the formation of the first generation of SMBHs and their mergers, surely a spectacular discovery.

\section{Acknowledgement:}
	
PLB wishes to thank his colleagues A.F. Barghouty, W. de Boer, S. Britzen, A. Brunthaler, A. Chieffi, R. Chini, P. Crumley, C. Dobrigkeit Chinellato, R. Engel, A. Fialkov, C. Galea, I. Gebauer, G. Gilmore, Gopal-Krishna, E. Haug, A. Haungs, M. Joyce, H. Kang, P.P. Kronberg, N. Menci, F. Munyaneza, G. P\u{a}v\u{a}la\c{s}, S. P\u{a}duroiu, J.P. Rachen, P. Rochus, D. Ryu, N. Sanchez, M. Sikora, G. Smoot, P. Sokolsky, M. Teshima, H. de Vega, and A. Zdziarski for intense discussions on these and related topics.

\end{document}